\documentclass[reqno,12pt]{article}
\usepackage{amsmath,amssymb,amsfonts,epsfig,graphicx,euscript}
\usepackage{graphics}
\newcommand{\bse}{\begin{subequations}}
\newcommand{\ese}{\end{subequations}}
\newcommand{\be}{\begin{equation}}
\newcommand{\ee}{\end{equation}}
\newcommand{\bea}{\begin{eqnarray}}
\newcommand{\eea}{\end{eqnarray}}
\newcommand{\ba}{\begin{array}}
\newcommand{\ea}{\end{array}}

\begin{document}

\begin{flushright}
\end{flushright}
\begin{flushright}
\end{flushright}
\hfill%
\begin{center}

{\LARGE {\sc Exact Solutions in Weyl-Cubed Gravity }}

\bigskip
{ Mohammad A. Ganjali
\footnote{ganjali@theory.ipm.ac.ir}} \\
{Department of Physics, Kharazmi University,\\P. O. Box
31979-37551, Tehran, Iran}
\\

\end{center}

\bigskip
\begin{center}
{\bf { Abstract}}\\
\end{center}
A unitary gravitational action up to third order of curvature in
which respects to the holographic $a-$theorem has been
constructed in \cite{myers}. In particular, its third order term
is just the Weyl-cubed term in four dimensions.

In this paper, we study this theory and find some its exact
classical solutions. We show that the theory admits conformally
flat, Lifshitz, Schr\"{o}dinger and also hyperscaling-violating
backgrounds as the solutions of equations of motion. Our analysis
has been done for the pure Weyl-cubed gravity, Einstein plus
Weyl-cubed term and gravity with matter.

\newpage
\section{Introduction}
Extended theories of gravity have attracted serious attempts
during the last decades. In fact, these theories have opened new
windows for studying some unsolved problems in Einstein gravity
such as quantization of metric field. Moreover, these extended
theories are good laboratories for testing some other aspects of
theoretical physics for example AdS/CFT \cite{Maldacena:1997re},
and shed also lights on other areas of physics i.e. condensed
matter physics.

Several approaches have been used to construct extended gravity
see, for example, \cite{lovel,Horava:2009uw}. One of the main
difference between these approaches is the method one applies to
obtain the action of the theory. For example, one may add new
terms in action up to Curvature-squared order. It was shown that
this extension of Einstein gravity is perturbatively
renormalizable \cite{Stelle:1976gc} but the theory contains
massive spin 2 ghostlike mode and massless graviton mode.

Among the several interesting prescriptions for finding the action
of extended gravity, the author of \cite{myers} were able to
calculate an action up to cubic order of curvature using a
concept which is called "the holographic a-theorem". The
resulting action is given by\footnote{Some other approaches have
been introduced for calculating the third order gravity see, for
example, \cite{Oliva:2010eb}}.
 \bea\label{action}
 I_G=\frac{1}{2l_p^{d-1}}\int\mathrm{d}^{d+1}x \, \sqrt{-g}\,
 \left(
\frac{d(d-1)}{L^2}\alpha+R+\mu L^2 {\cal X}+\beta L^4 {\cal
Z}\right),
 \eea
where ${\cal X}$ is the four-dimensional Euler density
\cite{lovel},
 \bea
{\cal X}_4=R_{abcd}R^{abcd}-4R_{ab}R^{ab}+R^2.
 \eea
${\cal Z}$ contains curvature-cubed interactions and it was
argued that a three-parameter family of unitary $R^3$
interactions would exist. The first is the cubic Lovelock
interaction which is proportional to the six-dimensional Euler
density ${\cal X}_6$
 \bea
 {\cal X}_6= \frac{1}{8}\,
\varepsilon_{abcdef}\,\varepsilon^{ghijkl}\,R_{ab}{}^{gh}\,
R_{cd}{}^{ij}\, R_{ef}{}^{kl}
 \eea
The second is the quasitopological interaction ${\cal Z}_{d+1}$,
 \bea
 {\cal Z}_{d+1}&=& R_a{}^c{}_b{}^{d} R_c{}^e{}_d{}^{f}R_e{}^a{}_f{}^{b}
 +
\frac{1}{(2d-1)(d-3)}\left(\frac{3(3d-5)}{8}R_{a b c d}R^{a b c d}
R \right.\nonumber\\
&&\;\;\;\;\;\;-\,3(d-1) R_{a b c d}R^{a b c}{}_{e}R^{d e}+
3(d-1)R_{a b c d}
R^{a c}R^{b d}\\
&&\;\;\;\;\;\;\left.+\,6(d-1)R_a{}^{b}R_b{}^{c}R_c{}^{a}-\frac{3(3d-1)}{2}
R_a{}^{b}R_b{}^{a}R +\frac{3(d-1)}{8}R^3\right)\,.\nonumber
 \eea
Two other basis are constructed from the Weyl tensor
 \bea
{\cal W}_1=W_{a\,\,b}^{\,\,c\,\,\,d}\, W_{c\,\,d}^{\,\,e\,\,\,f}\,
W_{e\,\,f}^{\,\,a\,\,\,b}\,,\;\;\;\;\;\; {\cal
W}_2=W_{ab}^{\,\,\,\,\,\,cd}\, W_{cd}^{\,\,\,\,\,\,ef}\,
W_{ef}^{\,\,\,\,\,\, ab}\,.
 \eea
These four basis are not all independent for $d\ge6$ due to the
following relation
 \bea
 {\cal Z}_{d+1}={\cal W}_1+\frac{3d^2-9d+4}{8(2d-1)(d-3)(d-4)}\left({\cal X}_6+8{\cal
 W}_1
 -4{\cal W}_2\right)\,.
 \eea
Thus, we have a three-parameter family of \textsl{unitary} cubic
interactions. For $d=5$, ${\cal Z}_6$ is not defined and so ${\cal
X}_6$, ${\cal W}_1$ and ${\cal W}_2$ are the bases. For $d<5$,
${\cal X}_6=0$ and by using Schouten identity, one has ${\cal
W}_1={\cal W}_2$ \cite{Sinha:2010pm}. Thus, in four dimension, we
have a one-parameter family of interactions ${\cal W }={\cal
W}_1={\cal W}_2$.

Investigation of classical solutions of this cubic theory of
gravity is of interest from several point of view. Indeed, first,
one may ask weather the well-known solutions of the Einstein
gravity are also the solutions of the cubic gravity or not.
Moreover, one may search for backgrounds in which are absent in
pure Einstein gravity \cite{myers, Deser:2003up, Oliva:2010eb,
Dehghani, Ganjali:2015cba}. Two classes of such solutions are
Lifshitz and Schr\"{o}dinger geometries. The Lifshitz background
is given by \cite{Kachru:2008yh}
 \bea\label{lifshitz1}
ds^2=-\frac{r^{2z}}{L^2}dt^2+\frac{L^2}{r^2}dr^2+\frac{r^2}{L^2}d\vec{x}^2,
 \eea
where $z$ is the dynamical exponent and exhibits space and time
scale differently
 \bea
t\mapsto \lambda^{z}t,\;\;\;\;r\mapsto
\lambda^{-1}r,\;\;\;\;\vec{x}\mapsto\lambda\vec{x}.
 \eea
Also, the line element of  Schr\"{o}dinger geometry is
\cite{Balasubramanian:2008dm}
 \bea\label{schro}
ds^2=L^2\left(-\frac{dt^2}{r^{2z}}+\frac{dr^2}{r^2}+\frac{2dtdx+dy^2}{r^2}\right),
 \eea
Its scaling property is as follows
 \bea
t\mapsto \lambda^{z}t,\;\;\;\;r\mapsto \lambda
r,\;\;\;\;x\mapsto\lambda^{2-z}x,\;\;\;\;\;y\mapsto \lambda
x,\;\;\;\;\;z\neq 1.
 \eea
Beside of having various interesting geometrical and physical
properties, these non-relativistic solutions have important
applications in condensed matter physics
\cite{Balasubramanian:2008dm, Charmousis:2010zz,
AyonBeato:2010tm}.

A generalization of these geometries are backgrounds which are
conformally related to these metrics \cite{Charmousis:2010zz}
 \bea\label{lifshitz}
d\tilde{s}^2=r^{-2\frac{\theta}{d-1}}ds^2.
 \eea
Here $\theta$ is called the hyperscaling violation exponent.
Recalling AdS/CFT, the non-zero $\theta$ means that in dual field
theory the hyperscaling violates and the entropy scales as
$T^{\frac{d-\theta-1}{z}}$.

In this paper our aim is to find such classical solutions for the
cubic theory in four dimensions. Recalling the previous
discussions, it is enough to consider ${\cal W}={\cal W}_1$ term
at cubic order in four dimensions. Note also that in four
dimensions the ${\cal X}_4$ is purely topological and does not
contribute in the equations of motion.

The paper is organized as follows. In the next section, we will
briefly introduce the cubic gravity. In section 2, we derive the
equations of motion for the Weyl-cubed gravity. After that, in
section 3, we study the possible classical solutions of the
theory. In particular, we will study the pure Weyl-cubed gravity,
Einstein plus Gauss-Bonnet plus Weyl-cubed gravity and gravity
with matter separately and will find some known solutions such as
conformally flat, Lifshitz and Schr\"{o}dinger metrics.

Let us set $L=1$ and define $\alpha=-\frac{\theta}{d-1}$ in the
rest of the paper.
\section{Equation of Motion and Solution}
Let us consider the action as
 \bea\label{actio}
I=I_G+I_M
 \eea
where $I_G$ denotes the gravitational part of the action in which
its cubic term ${\cal Z}$ contains only\footnote{In this section,
we present the action and equations of motion for general $d+1$
dimensions but we will obtain the solutions in 4 dimension. Thus
we neglect the ${\cal Z}_{d+1}$ term in (\ref{actionn}).} ${\cal
W}={\cal W}_1$ and $I_M$ is the matter part of the action
 \bea\label{actionn}
I=-\frac{1}{2l_p^{d-1}}\int\mathrm{d}^{d+1}x\sqrt{-g}\left(d(d-1)\Lambda
+R+\mu{\cal X}_4 +\beta{\cal W}\right)+I_M.
 \eea
 Varying the action with respect to metric gives us the equations of
 motion as follows
 \bea
G_{\mu\nu}-\frac{1}{2}d(d+1)\Lambda g_{\mu\nu}+\mu
E_{\mu\nu}+\beta {\cal G}_{\mu\nu}=2\frac{l_p^{d-1}}{\sqrt{-g}}
\frac{\delta I_M}{\delta g_{\mu\nu}},
 \eea
where $G_{\mu\nu}$ is the Einstein tensor and $E_{\mu\nu}$ and
${\cal G}_{\mu\nu}$ are given by

 \bea\label{E}
E_{\mu\nu}&=&-\frac{1}{2}{\cal
X}_4g_{\mu\nu}+2R_{\mu}^{\;\;\theta\sigma\gamma}R_{\nu\theta\sigma\gamma}-4R_{\mu\theta\nu\gamma}R^{\theta\gamma}-4R_{\mu\theta}R_{\nu}^{\;\theta}
+2RR_{\mu\nu},\\
 \label{G}
{\cal
G}_{\mu\nu}&=&-\frac{1}{2}W_{\rho\theta\sigma\gamma}\digamma^{\rho\theta\sigma\gamma}g_{\mu\nu}
+6W_{\mu\theta\sigma\gamma}\digamma_{\nu}^{\;\;\theta\sigma\gamma}\\
&&-3R_{\nu\theta\sigma\gamma}\digamma^{\;\;\theta\sigma\gamma}_{\mu}-\frac{12}{d(d-1)}(dR_{\theta\sigma}-Rg_{\theta\sigma})
\digamma_{\mu\;\;\;\;\nu}^{\;\;\theta\sigma}
+\frac{6}{d(d-1)}R_{\mu\nu}\digamma_{\rho\theta}^{\;\;\;\;\rho\theta}\cr
&&-6\nabla_{\theta}\nabla_{\sigma}\digamma_{\mu\;\;\;\;\nu}^{\;\;\theta\sigma}
-\frac{6}{d-1}\nabla_{\rho}\nabla_{\gamma}\digamma^{\rho\theta\;\;\gamma}_{\;\;\;\;\theta}g_{\mu\nu}
-\frac{12}{d-1}\nabla_{\rho}\nabla_{(\mu}\digamma_{\;\;\;\;\theta\nu)}^{\rho\theta}
\cr
&&-\frac{6}{d(d-1)}\nabla_{\mu}\nabla_{\nu}\digamma_{\rho\theta}^{\;\;\;\;\rho\theta}
-\frac{6}{d-1}\nabla^2\digamma_{\mu\theta\;\;\nu}^{\;\;\;\;\theta}
+\frac{6}{d(d-1)}\nabla^2\digamma_{\rho\theta}^{\;\;\;\;\rho\theta}g_{\mu\nu}\nonumber
 \eea
and it was used the following definition in (\ref{G})
 \bea
\digamma_{\rho\theta\sigma\gamma}=W_{\rho\theta}^{\;\;\;\;\epsilon\kappa}W_{\epsilon\kappa\sigma\gamma}.
 \eea
It is notable that the above expression for ${\cal G}_{\mu\nu}$
can be simplified in four dimensions using the following relations
\cite{Lu:2012xu}
 \bea
W_{\rho\theta\sigma\gamma}W^{\rho\theta\sigma\gamma}&=&R_{\rho\theta\sigma\gamma}R^{\rho\theta\sigma\gamma}
-2W_{\rho\theta}W^{\rho\theta}+\frac{1}{3}R^2\\
\digamma_{\mu\theta\;\;\nu}^{\;\;\;\;\theta}&=&-W_{\mu\theta\sigma\gamma}W_{\nu}^{\;\;\theta\sigma\gamma}=
-\frac{1}{4}W_{\rho\theta\sigma\gamma}W^{\rho\theta\sigma\gamma}g_{\mu\nu}.
 \eea
 \section{Solution}
In this section, we would like to find the solutions of the
equations of motion. We will first consider the pure Weyl-cubed
gravity and try to obtain the vacuum solutions i.e. the solutions
of ${\cal G}_{\mu\nu}=0$. After that, we add the usual
Einstein-Hilbert term to the action. At the end, we will consider
the full action (\ref{actio}) with a particular form of matter
action and discuss about the possible solutions of the theory.
\subsection{Pure Weyl-Cubed Gravity}
First of all, we would like to investigate the pure Weyl-cubed
gravity. It is notable that the coefficient of ${\cal W }$ term
in the action (\ref{actionn}) has dimension $(mass)^{d-5}$ which
means that this term is \textit{super-renormalizable} in $d=3,4$
dimensions, \textit{renormalizable} in $d=5$ dimension and is
\textit{non-renormalizable} for $d>5$. In particular, in four
dimension the pure Weyl-cubed gravity is super-renomalizable.

For obtaining the classical backgrounds of this theory we should
solve
 \bea\label{puree}
 {\cal G}_{\mu\nu}=0
 \eea
which is a set of fourth order differential equations and it is
not obvious how they can be integrated in generality. However,
with a glimpse to ${\cal G}_{\mu\nu}$, one can realize two
classes of solutions of (\ref{puree}). The first is backgrounds
with $W_{\mu\nu\rho\sigma}=0$ (conformally flat spaces) and the
second is backgrounds with $\digamma_{\mu\nu\rho\sigma}=0$. Note
also that any conformally related spaces to these two classes of
solutions do also solve the (\ref{puree}).
\subsubsection{Solutions with $W_{\rho\sigma\theta\gamma}=0$}
It is easy to see that the equation
$W_{\rho\sigma\theta\gamma}=0$ can be solved if one considers
$$R_{\rho\sigma\theta\gamma}=2\Lambda
(g_{\rho\theta}g_{\sigma\gamma}-g_{\rho\gamma}g_{\sigma\theta}).$$
As a result, geometries which are locally flat, dS or AdS are
solutions of pure Weyl-cube gravity \cite{Deser:2003up,
Oliva:2010eb}. Here the parameter $\Lambda$ is an arbitrary
constant and so, the limiting case $\Lambda=0$ gives us the usual
Schwarzschild solution.

In order to be more explicit let us consider, for simplicity, the
static spherically symmetric background as
 \bea\label{sphere}
ds^2=-r^{2}f(r)dt^2+\frac{dr^2}{r^2h(r)}+r^2d\Omega_k^2,
 \eea
where $d\Omega^2_{k}$ is the line element on unit 2-sphere,
hyperbolic and flat spaces with $k=1,-1,0$ respectively. Using
the above metric and doing a straightforward calculation, one can
see that in order to have conformally flat space it is enough
that the $f(r)$ and $h(r)$ satisfy the following equation
 \bea\label{h}
\frac{f''}{f}-2(\frac{f'}{f})^2+\frac{2}{r}\frac{f'}{f}+2\frac{f'}{f}\frac{h'}{h}-\frac{2k}{r^4h}=0,
 \eea
where the prime stands for derivative with respect to $r$
coordinate. Then, one can integrate (\ref{h}) and obtain $h(r)$ in
terms of $f(r)$ as follows
 \bea\label{hh}
h(r)=\frac{(4kf(r)+c)f(r)}{(r^2f'(r))^2},
 \eea
where $c$ is a constant of integration. We see that there is
degeneracy in the solution space of the theory and that there is
no corresponding Birkhoff theorem in pure Weyl-cubed gravity.

As some examples, for the case where $c=k=0$ and $f(r)=h(r)=1$ one
recovers the AdS space\footnote{Note also that, in (\ref{h}) when
$f(r)=const$, then $k=0$ and so $h(r)$ is free and the solution
is degenerate.}. Moreover, setting $c=0$ and
$f(r)=(1-\frac{M}{r})^{\alpha}$ then we have
$h(r)=\frac{4k}{(\alpha M)^2}(1-\frac{M}{r})^2$ which exhibits a
black hole solution with horizon at $r=M$. Furthermore, setting
again $c=0$ and considering $f(r)=\frac{4}{(c_1+c_2r)^2}$ gives
us a solution with $R=0$ provided that $k\neq 0$. As concluding
example, let us consider $f(r)=r^{2z-2}h(r)$. Then, we obtain
 \bea\label{fsymm}
f(r)=\frac{c_1}{r^z}+\frac{c_2}{r^{2z-2}}+\frac{k}{(z-2)^2r^2},\;\;\;\;\;\;z\neq
2 ,
 \eea
which is asymptotically Lifschitz space only for $z=\{0,1\}$.

At the end, it is important to note that since time dependent FRW
backgrounds are conformally flat thus they are vacuum solutions of
the pure Weyl-cubed gravity.
\subsubsection{Solutions with $\digamma_{\rho\sigma\theta\gamma}=0$}
In this section we want to find solutions of
 \bea\label{equu}
\digamma_{\rho\sigma\theta\gamma}=0.
 \eea
Since for any static spherically symmetric space with
$\digamma_{\rho\sigma\theta\gamma}=0$ then we have
$W_{\rho\sigma\theta\gamma}=0$ and vice versa, we search for some
metrics in which $W_{\rho\sigma\theta\gamma}\neq 0$ i.e.
backgrounds with different symmetry. In particular, we examine the
metric ansatz with Shr\"{o}dinger or asymptotically
Shr\"{o}dinger symmetry. For this aim, we consider a generalized
form of (\ref{schro}) as follows
 \bea\label{ansatzs}
ds^2=-\frac{dt^2}{r^{2z}f(r)}+\frac{dr^2}{r^2h(r)}+2\frac{dtdx}{r^2j(r)}+\frac{dy^2}{r^2},
 \eea
Then, the equation (\ref{equu}) reads as
 \bea
\frac{j''}{j}-(\frac{j'}{j})^2+\frac{1}{2}\frac{j'}{j}\frac{h'}{h}=0,
 \eea
 and can be solved as
 \bea\label{solschro}
h(r)=c(\frac{j(r)}{j'(r)})^2.
 \eea
Observe that with the solution (\ref{solschro}) the Weyl tensor is
in general nonzero\footnote{One can show that the solution
(\ref{solschro}) with $j(r)=\frac{1}{c_1+c_2r^{2-2z}f(r)}$
implies that
$\digamma_{\rho\sigma\theta\gamma}=W_{\rho\sigma\theta\gamma}=0$.}.
Here, we have a two-fold degeneracy in the solution. Both the
$f(r)$ and $j(r)$ are undetermined. For $j(r)=1$ then
(\ref{equu}) automatically is satisfied and by $h(r)=j(r)=1$ we
recover the Schr\"{o}dinger geometry (\ref{schro}).

It is notable that since the pp-wave background with the line
element
  \bea\label{ansatzpp}
ds^2=\frac{dr^2}{r^2h(r)}+r^{2}f(r)dx^2+2r^2j(r)dtdx+r^2dy^2,
 \eea
is conformally related to Schr\"{o}dinger metric, it is also the
vacuum solution of pure Weyl-cubed gravity.
\subsection{Einstein+Weyl-Cubed Gravity}
In this section, we would like to discuss pure gravity theory in
four dimensions in which includes both the Einstein term,
Weyl-cubed term and also cosmological constant in the action. The
equations of motion are given by
 \bea\label{eew}
G_{\mu\nu}-\frac{1}{2}d(d+1)\Lambda g_{\mu\nu}+\beta {\cal
G}_{\mu\nu}=0.
 \eea
Note that one may also add the curvature squared term, i.e.
Gauss-Bonnet action, but in four dimensions such term is purely
topological and does not contribute in the equations of motion.

In searching for classical solutions of this theory, we can
classify them into two classes, those that ${\cal G}_{\mu\nu}=0$
and those that ${\cal G}_{\mu\nu}\neq 0$. For the first class, we
have classified some possible solutions of ${\cal G}_{\mu\nu}=0$
in the previous sections. However, those geometries should also
obey the usual Einstein equation $G_{\mu\nu}=3\Lambda
g_{\mu\nu}$. From the Einstein equation, we have
$R_{\mu\nu}=-6\Lambda g_{\mu\nu}$. Therefore, backgrounds that
saturate $R_{\rho\sigma\theta\gamma}=2\Lambda
(g_{\rho\theta}g_{\sigma\gamma}-g_{\rho\gamma}g_{\sigma\theta})$
are solutions of the full Einstein+Weyl-cubed gravity. As an
example, by setting $\Lambda=0$, we find the following metric
 \bea
ds^2=-r^{2z-2}f(r)dt^2+\frac{dr^2}{r^4f(r)}+d\Omega_k^2,
 \eea
where $f(r)$ was given in (\ref{fsymm}).

Now, we focus on the second class where ${\cal G}_{\mu\nu}\neq
0$. We also consider the following hyper-scaling violating static
spherically symmetric metric
 \bea\label{gsphere}
ds^2=r^{2\alpha}\left(-r^{2z}f(r)dt^2+\frac{dr^2}{r^2f(r)}+r^2d\Omega_k^2\right),
 \eea
where $\alpha$ exhibits the hyper-scaling violation and $z$ is
dynamical exponent. By these assumptions, however, solving the
equation (\ref{eew}) with ${\cal G}_{\mu\nu}\neq 0$ is a very
hard task. However, one can show that the solutions with
Lifschitz asymptotic do exist only for \cite{Ganjali:2015cba}
 \bea
\alpha=0,-1.
 \eea
Indeed, using the ansatz $f(r)=1-\frac{s}{r^2}$, where $s$ is a
constant, one finds the following solution(with ${\cal
G}_{\mu\nu}\neq 0$)
 \bea
&&a)\alpha=0\;\;\;\;\;\;\;\;s=0,\;\;\;\;k=0,\;\;\;\;\Lambda\neq
0,\;\;\;\;z\neq \{0,1,4\}\hspace{3cm}\nonumber\\
&&\hspace{2.3cm} \Lambda=\frac{18 + 7 z + 4 z^2 - 2 z^3}{9(4 -
z)},\;\;\;\beta=-\frac{3}{2 z^2 (4 - 5 z + z^2)},\nonumber\\
&&b)\alpha=-1\;\;\;\;\;s\neq 0,\;\;\;\;k=0,\;\;\;\;\Lambda\neq
0\;\;\;\;\;\;z=\{0,1\}\nonumber\\
&&\hspace{2.3cm}
s=-\frac{(2z+1)^2}{(z+1)}\Lambda,\;\;\;\;\;\;\;\;\;\;\beta=\frac{27}{4(z-2)^6}\frac{\Lambda}{s^3}.\nonumber
 \eea
Observe that in the case (b) we have a black hole solution with
flat horizon. We can also compute the scalar invariant
$W_{\rho\theta\sigma\gamma}W^{\rho\theta\sigma\gamma}$ and obtain
$$W_{\rho\theta\sigma\gamma}W^{\rho\theta\sigma\gamma}=\frac{4}{3}(z-2)^4s^2,$$
which indicates that this black hole has not curvature
singularity. It is noticeable that for generating
hyperscaling-violating solution in usual Einstein gravity one
should consider the coupling between matter and the background
metric \cite{Tarrio:2011de} but, as we have already showed, such
solution can be generated in the pure Weyl-cubed and
Einstein+Weyl-cubed theories.
\subsection{Cubic Gravity with Matter}
In this section, we first briefly introduce the
Einstein-dilaton-Maxwell theory where has been constructed in
\cite{Alishahiha:2012qu} and its nonlinear generalization has
been studied in \cite{Dehghani:2015gza}\footnote{We will present
the results of this section for a general $d+1$ dimensional
space.}. The Lagrangian of the matter part of this theory ${\cal
L}_M$ in $d+1$ dimensions may be written as\footnote{Here, we
write a generalized version of the action used in
\cite{Alishahiha:2012qu, Dehghani:2015gza}}
 \bea\label{actionmatter}
{\cal L}_M= -\frac{1}{2}(\partial\phi)^2+V(\phi)+
\sum_{i=1}^2\left(-\frac{1}{4}e^{\lambda_i\phi}F_i^2+\frac{1}{4}e^{\epsilon_i\phi}(-H_i^2)^{s_i}\right),
 \eea
where $\lambda_1, \lambda_2, \epsilon_1, \epsilon_2$ are some free
parameters. The field strength are defined as usual for linear
gauge fields $A_{i\;\mu}$ as
$F_{i\;\mu\nu}=\partial_{\mu}A_{i\;\nu}-\partial_{\nu}A_{i\;\mu}$
and for nonlinear fields $B_{i\;\mu}$ as
$H_{i\;\mu\nu}=\partial_{\mu}B_{i\;\nu}-\partial_{\nu}B_{i\;\mu}$.
Two constants $s_i$-s demonstrate the nonlinearity of the action.
We will see that the presence of nonlinear field is necessary in
order to solve all the equations of motion. Motivated by string
theory it was considered an exponential potential for dilaton
$V=V_0e^{\gamma\phi}$ where $V_0$ is a free parameter
\cite{Alishahiha:2012qu}.

Varying the action with respect to $ A_{i\;\mu}, B_{i\;\mu}, \phi$
and $g_{\mu\nu}$ give us the following equations of motion
 \bea
\label{equations2} &&\nabla_{\mu}\left(\sqrt{-g}e^{\lambda_i\phi}
F_i^{\mu\nu}\right)=0,\\
\label{equations4}
&&\nabla_{\mu}\left(s_i\sqrt{-g}e^{\epsilon_i\phi}
(-H^2)^{s_i-1}H_i^{\mu\nu}\right)=0,\\\label{equations1}
&&\nabla^2\phi+\frac{dV(\phi)}{d\phi}-\frac{1}{4}\sum_{i}\left(\lambda_i
e^{\lambda_i\phi}F_i^2-\epsilon_i
e^{\epsilon_i\phi}(-H_i^2)^{s_i}\right)=0,\\
 \label{equations3}
&&R_{\mu\nu}+\beta\left({\cal G}_{\mu\nu}-\frac{{\cal G
}}{d-1}g_{\mu\nu}\right)+\\
&&-\frac{1}{2}\partial_{\mu}\phi\partial_{\nu}\phi+\frac{V(\phi)}{d-1}g_{\mu\nu}-\frac{1}{2}\sum_i
e^{\lambda_i\phi}\left(F^{\rho}_{i\;\mu}F_{i\;\rho\nu}-\frac{F_i^2}{2d-2}g_{\mu\nu}\right)+\cr
&&-\frac{1}{2}\sum_i
e^{\epsilon_i\phi}\left(s_i(-H_i^2)^{s_i-1}H^{\rho}_{i\;\mu}H_{i\;\rho\nu}
-(\frac{2s_i-1}{2d-2})(-H_i^2)^{s_i}g_{\mu\nu}\right)=0.\nonumber
 \eea
where ${\cal G}=G^{\mu}_{\;\;\mu}$. We suppose the ansatz
(\ref{gsphere}) with $k=0$, and following ansatz for the scalar
and gauge fields
 \bea\label{ansatz1}
ds^2&=&r^{2\alpha}\left(-r^{2z}f(r)dt^2+\frac{dr^2}{r^2f(r)}+r^2d\vec{x}^2\right),\cr
\phi&=&\phi(r),\;\;\;\;\;\;F_{i\;rt}\neq
0,\;\;\;\;\;\;H_{i\;rt}\neq 0.
 \eea
Here, we again face with a set of very difficult differential
equations. Therefore, for simplicity, we search for solutions in
which $$W_{\rho\theta\sigma\gamma}=0.$$ To proceed, first, recall
that the conformally flat metrics have been given in
(\ref{fsymm}). Then, since ${\cal G}_{\mu\nu}=0$, we deal with
the usual Einstein-dilaton-Maxwell theory. The classical solutions
of this theory have been obtained in \cite{Alishahiha:2012qu,
Dehghani:2015gza}. Let us present the results for the more
general equations (\ref{equations2}-\ref{equations3}).

Using the ansatz (\ref{ansatz1}) and Maxwell equation one obtains
 \bea
F_{i\;rt}&=&\rho_ie^{-\lambda_i\phi}r^{\alpha(3-d)+z-d}\\
H_{i\;rt}&=&\varrho_ie^{-\frac{\epsilon_i\phi}{2s_i-1}}r^{\frac{(2s_i-1)(z+2\alpha-1)-(d-1)(\alpha+1)}{2s_i-1}}.
 \eea
Then, by focusing on $R^t_{t}-R^r_r$ components of equation
(\ref{equations3}) and noting that for the metric (\ref{ansatz1})
we have $(R^t_{t}-R^r_r)\thicksim r^{-2\alpha}f(r)$, we are able
to find the scalar field as
 \bea
e^{\phi}=e^{\phi_0}r^{\sqrt{2(d-1)(\alpha+1)(\alpha+z-1)}}=e^{\phi_0}r^{\Gamma}.
 \eea
It is clear that in order to have a well defined solution we have
to consider $(\alpha+1)(\alpha+z-1)\geq 0$. In fact, this
condition is a result of null energy condition. Indeed, one
utilizes the $xx$ components of (\ref{equations3}) and integrate
it to find $f(r)$ as follows
 \bea\label{ansatz2}
f(r)=&-&mr^{-(d-1)\alpha-z-d+1}\\
&+&\frac{V_0e^{\gamma\phi_0}r^{\gamma\Gamma+2\alpha}}{(d-1)(\alpha+1)(\gamma\Gamma+\alpha(d+1)+z+d-1)}\nonumber\\
&-&\sum_{i=1}^2\frac{\rho_i^2e^{-\lambda_i\phi_0}r^{-2\alpha(d-2)-\Gamma\lambda_i-2d+2}}{2(d-1)(\alpha+1)(\alpha(3-d)+z-d+1-\Gamma\lambda_i)}\nonumber\\
&-&\sum_{i=1}^2
\frac{(2\varrho_i^2)^{s_i}(2s_i-1)^2e^{-\frac{\epsilon_i\phi_0}{2s_i-1}}r^{\frac{\left(-2+(6-2d)s_i\right)\alpha-2s_i\left(d-1\right)
-\Gamma\epsilon_i}{2s_i-1}}}{4(d-1)(\alpha+1)\left(\alpha(4s_i-d-1)+(2s_i-1)z-d+1-\Gamma\epsilon_i\right)},\nonumber
 \eea
where $m$ is a constant. At the next step, we focus on the
equation of motion of the scalar field which can be written as
 \bea
 \left(\frac{4\Gamma}{(d-1)(\alpha+1)}+4\gamma\right)V_0e^{\gamma\phi}&=&
 \sum_{i=1}^{2}e^{\lambda_i\phi}F_i^2\left(\lambda_i-\frac{\Gamma}{(d-1)(\alpha+1)}\right)\\
 &-&\sum_{i=1}^{2}e^{\epsilon_i\phi}(-H_i^2)^{s_i}\left(\epsilon_i-\frac{(2s_i-1)\Gamma}{(d-1)(\alpha+1)}\right),\nonumber
 \eea
The left and right hand sights of this equation can be equal to
each other if we choose the parameters $\lambda_1, \lambda_2,
\epsilon_i$ and $\rho_1$ as follows
 \bea\label{la1}
&&\lambda_1=-\frac{\gamma\Gamma+2(d-1)(\alpha+1)}{\Gamma},\;\;\;\;\;\;\;\;\;\;\;\lambda_2=\frac{\Gamma}{(d-1)(\alpha+1)}\nonumber\\
&&\epsilon_i=\frac{(2s_i-1)\Gamma}{(d-1)(\alpha+1)},\nonumber\\
&&\rho_1^2=2\frac{\Gamma+(d-1)(\alpha+1)\gamma}{\Gamma-(d-1)(\alpha+1)\lambda_1}V_0e^{(\gamma+\lambda_1)\phi_0}.
 \eea
With these relations at hand, we have several possibilities in
which the (\ref{fsymm}) coincides with (\ref{ansatz1}). Note that
such coincidence should respect not only the remaining equations
of motion but also should satisfy the null energy
condition\footnote{In fact, the null energy condition usually rule
out some other possibilities for which (\ref{fsymm}) coincides
with (\ref{ansatz1}).}. However, for presenting a consistent
solution, we first set
 \bea\label{gamma}
 -(d-1)\alpha-z-d+1=2-2z,\;\;\;\;\;\;\;\;\;\gamma\Gamma+2\alpha=-z,
 \eea
which imply that
 \bea\label{alga}
\alpha=\frac{z-d-1}{d-1},\;\;\;\;\;\;\;\;\;\;\;\;\;\;\;\;\Gamma=\sqrt{\frac{2d}{d-1}}(z-2).
 \eea
It is clear that $(\alpha+1)(\alpha+z-1)\geq 0$ for generic value
of $z$. We can also see that with (\ref{gamma}) the power of $r$
in the third line of (\ref{ansatz2}) for the first linear gauge
field $A_{1\;\mu}$ is $\gamma\Gamma+2\alpha=-z$. Moreover, using
(\ref{alga}) and plugging $\lambda_2$ from (\ref{la1}) back into
the solution (\ref{ansatz2}) and equating the resultant with $-z$
or $2-2z$ implies that for the second linear gauge field we
should set $z=2$. Accordingly, we obtain
$$z=2,\;\;\;\;\alpha=-1,\;\;\;\;\Gamma=\lambda_2=0,\;\;\;\;\phi=const,\;\;\;\;F_{i\;rt}\sim
\frac{1}{r},$$ and the metric reduces to Minkowski metric.
Therefore, let us set the free parameter $\rho_2=0$. So, we have
only one linear $U(1)$ gauge field in the theory.

It is worth mentioning that if we choose
$\gamma\Gamma+2\alpha=2-2z$ then $\rho_1=0$ so this option for
the power of the $r$ in the second line of (\ref{ansatz2}) should
be excluded.

In the fourth line of (\ref{ansatz2}) we have two terms in which
the power of $r$ equals to
$\left(-2+(6-2d)s_i\right)\alpha-2s_i\left(d-1\right)
-\Gamma\epsilon_i$ and two free parameters $\varrho_i$-s. Up to
now, we are free to set the powers of these terms either $-z$ or
$2-2z$. But, we shall soon see that in order to satisfy the
remaining equations of motion, i.e. $tt$ and $rr$ components of
(\ref{equations3}), one should set the power of the at least one
of these terms to $-z$. Doing this for the first nonlinear gauge
field $B_{1\;\mu}$, one obtains
 \bea
s_1=\frac{1}{4},
 \eea
For the second nonlinear gauge field we are yet free to set the
power of $r$ to $-z$ or $2-2z$. However, one finds that for the
later case $\alpha=-1$ which is not a valid value. So the powers
of $r$ in the fourth term of (\ref{ansatz2}) are also equal to
$-z$ and we have $s_1=s_2=\frac{1}{4}$.

The remaining equation of motion i.e. $tt$(or $rr$) component of
(\ref{equations3}) helps us to find a relation between the charges
of nonlinear gauge fields and $V_0$ and $\phi_0$ as follows
 \bea
 \sum_{i=1}^2q_i^{\frac{1}{2}}e^{2\epsilon_i\phi_0}=-\frac{2^{\frac{15}{4}}}{3}V_0e^{\gamma\phi_0}
 \eea
Plugging all the results back into the (\ref{ansatz2}) gives us
the final form of the $f(r)$ as
 \bea
f(r)=-mr^{2-2z}.
 \eea
 As we see, the contributions from the second, third and forth
 lines of (\ref{ansatz2}) cancel each other and we have a metric
 with one free parameter $m$.
\section{Conclusion}
In this paper, we have studied the classical solutions of a
gravitational theory where has been constructed in \cite{myers}.
The action of this theory is of order three of curvature tensor.
Such cubic action has been obtained by using some simple
"Holographic c/a-theorem" in arbitrary dimensions. In particular,
it involves only Weyl-cubed term at third order in four
dimensions.

We have analyzed the solutions of the pure Weyl-cubed gravity,
Einstein plus Weyl-cubed term and gravity with matter separately.
In pure Weyl-cubed gravity, we classified the solutions to those
that $W_{\mu\nu\lambda\theta}=0$ and those that
$W_{\mu\nu\lambda\theta}W^{\mu\nu\lambda\theta}=0$. The former
case involves the conformally flat metrics. For the later case,
we obtained the Schr\"{o}dinger space as a solution of pure cubic
gravity in for dimensions.

After that, we have considered the full
Einstein+Gauss-Bonnet+Weyl-cubed gravity. Although solving the
equations of motion in this case is very hard but we found a
possible solution for this theory. In particular, we supposed a
background with non-zero hyper scaling violating exponent
$-2\frac{\theta}{d-1}$ and dynamical exponent $z$ and found that
the solutions do exist only for $\theta=0,2$ in four dimensions.
When $\theta=0$, we have Lifshitz solution for any value of $z$
except of $z=\{0,4\}$. Moreover, when $\theta=2$, we have black
hole solution for $z=\{0,1\}$. The above solutions exist with
certain constrains on parameters of the theory.

At the end, we added the matter to the theory. The matter part
includes a scalar field, linear gauge fields and nonlinear gauge
fields \cite{Alishahiha:2012qu, Dehghani:2015gza}. We found the
conformally flat solutions of this theory i.e.
$W_{\mu\nu\lambda\theta}=0$.
\section{Acknowledgment}
Mohammad A. Ganjali would like to thanks the Kharazmi university
for supporting the paper with a grant.

\end{document}